\documentclass[sigconf]{acmart}

\usepackage[super]{nth}
\usepackage{amsmath}
\usepackage{amsthm}
\usepackage{mathtools}
\usepackage{bm}
\usepackage{array}
\usepackage{makecell}
\usepackage{multirow}
\usepackage{tabularx}
\usepackage{graphicx}
\usepackage{caption}
\usepackage{subcaption}
\usepackage{enumitem}
\usepackage{url}
\usepackage{xcolor}
\usepackage{enumitem}
\usepackage{bbm}
\usepackage{float}
\usepackage[flushleft]{threeparttable}
\usepackage[ruled,vlined,linesnumbered]{algorithm2e}
\SetArgSty{textnormal}
\usepackage{wrapfig,lipsum,booktabs}
\usepackage{subcaption}
\usepackage{hyperref}
\usepackage{makecell}

\DeclareMathOperator*{\argmax}{arg\,max}

\usepackage{xcolor,pifont}
\definecolor{cycle2}{RGB}{106, 191, 0}
\definecolor{cycle3}{RGB}{191, 0, 0}
\newcommand{\cmark}{\textcolor{cycle2}{\ding{52}}}
\newcommand{\xmark}{\textcolor{cycle3}{\ding{56}}}

\usepackage{graphicx} 

\theoremstyle{definition}

\newcounter{todocounter}
\newtoggle{comments}
 \toggletrue{comments} 
\NewEnviron{doweprint}[1]{%
  \iftoggle{#1}{\BODY}{}%
}

\AtBeginDocument{%
  \providecommand\BibTeX{{%
    \normalfont B\kern-0.5em{\scshape i\kern-0.25em b}\kern-0.8em\TeX}}}

\copyrightyear{2024}
\acmYear{2024}
\setcopyright{acmlicensed}
\acmConference[WWW '24 Companion] {Companion Proceedings of the ACM Web Conference 2024}{May 13--17, 2024}{Singapore, Singapore.}
\acmBooktitle{Companion Proceedings of the ACM Web Conference 2024 (WWW '24 Companion), May 13--17, 2024, Singapore, Singapore}
\acmISBN{979-8-4007-0172-6/24/05}
\acmDOI{10.1145/XXXXXX.XXXXXX}

\begin{document}

\title[Knowledge Graph-based Session Recommendation with Adaptive Propagation]{Knowledge Graph-based Session Recommendation with Session-Adaptive Propagation}

\author{Yu Wang}
\email{yu.wang.1@vanderbilt.edu}
\affiliation{%
  \institution{Vanderbilt University}
  \city{Nashville}
  \state{TN}
  \country{USA}
}

\author{Amin Javari}
\email{amin_javari@homedepot.com}
\affiliation{%
  \institution{The Home Depot}
  \city{Atlanta}
  \state{GA}
  \country{USA}
}

\author{Janani Balaji}
\email{janani_balaji@homedepot.com}
\affiliation{%
  \institution{The Home Depot}
  \city{Atlanta}
  \state{GA}
  \country{USA}
}

\author{Walid Shalaby}
\email{walid_shalaby@homedepot.com}
\affiliation{%
  \institution{The Home Depot}
  \city{Atlanta}
  \state{GA}
  \country{USA}
}

\author{Tyler Derr}
\email{tyler.derr@vanderbilt.edu}
\affiliation{%
  \institution{Vanderbilt University}
  \city{Nashville}
  \state{TN}
  \country{USA}
}

\author{Xiquan Cui}
\email{xiquan_cui@homedepot.com}
\affiliation{%
  \institution{The Home Depot}
  \city{Atlanta}
  \state{GA}
  \country{USA}
}

\begin{abstract}
Session-based recommender systems (SBRSs) predict users' next interacted items based on their historical activities. While most SBRSs capture purchasing intentions locally within each session, capturing items' global information across different sessions is crucial in characterizing their general properties. Previous works capture this cross-session information by constructing graphs and incorporating neighbor information. However, this incorporation cannot vary adaptively according to the unique intention of each session, and the constructed graphs consist of only one type of user-item interaction. To address these limitations, we propose knowledge graph-based session recommendation with session-adaptive propagation. Specifically, we build a knowledge graph by connecting items with multi-typed edges to characterize various user-item interactions. Then, we adaptively aggregate items' neighbor information considering user intention within the learned session. Experimental results demonstrate that equipping our constructed knowledge graph and session-adaptive propagation enhances session recommendation backbones by 10\%-20\%. Moreover, we provide an industrial case study showing our proposed framework achieves 2\% performance boost over an existing well-deployed model at The Home Depot e-platform.

\vspace{-1ex}
\end{abstract}

\ccsdesc[500]{Computing methodologies~Machine learning\vspace{-1.5ex}}


\keywords{Session Recommendation, Knowledge Graph, Adaptive Propagation}

\maketitle

\section{Introduction}\label{sec-introduction}
\begin{figure}[t]
     \centering
     \includegraphics[width=0.48\textwidth]{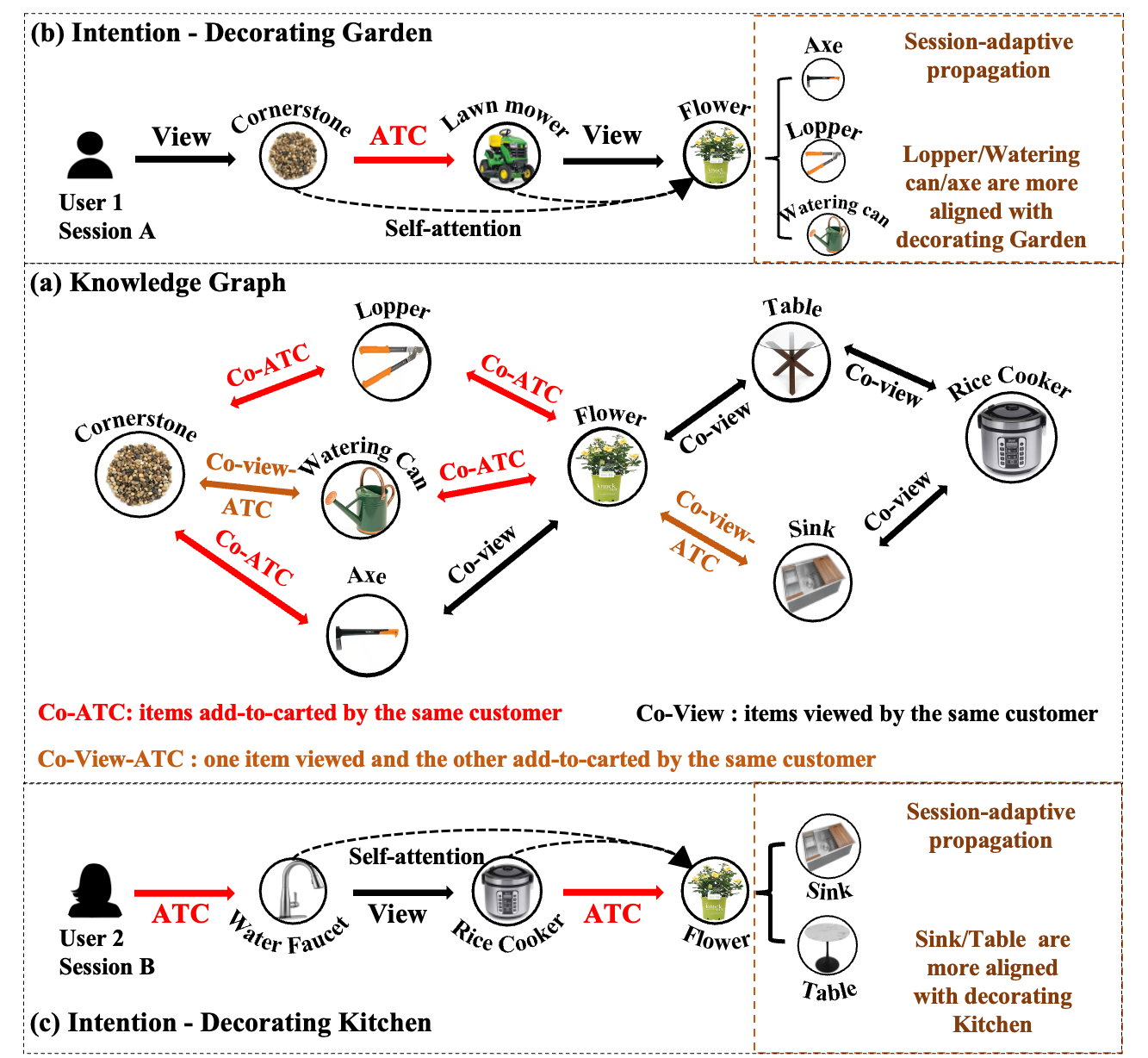}
     \vspace{-4ex}
     \caption{Since the user in Session A (b) intends to decorate the garden, while the user in Session B (c) intends to decorate the kitchen, the corresponding neighbors from the knowledge graph (a) are different for the same flower. By our proposed session-adaptive propagation, the flower aggregates more information from the lopper/watering can in (b) while more information from the sink/table in (c).}
     \label{fig-app_verify}
     \vspace{-3.5ex}
\end{figure}

Transformer-based models have shown state-of-the-art performance for session-based recommender systems (SBRSs) by leveraging their attention mechanisms and deep learning capabilities ~\cite{hidasi2015session, xu2019recurrent}. While transformers are adept at capturing local session data and individual item preferences~\cite{kang2018self, de2021transformers4rec}, they are limited in their ability to capture global transitional patterns among items~\cite{zhang2021learning, huang2021graph, qiu2020exploiting}. This limitation has sparked research into hybrid graph-based SBRSs that combine the strengths of transformers with Graph Neural Networks (GNNs) to capture both local and global dependencies~\cite{huang2021graph, zhang2021learning, qiu2020exploiting, wu2019session}. For example, in Figure~\ref{fig-app_verify}(b), borrowing the transitional information between flower and lopper in Figure~\ref{fig-app_verify}(a) helps characterize the intent of session A as decorating the garden and hence increases the probability of predicting the next item to be garden-related, e.g., watering can.

However, the existing hybrid models face two significant challenges. Firstly, the lack of full connectivity between GNN and transformer models limits their ability to capture the session dynamics~\cite{xia2021knowledge, pan2021graph, wang2020global}. While GNNs excel at capturing item-item relationships, they often struggle to incorporate the broader session context, and the learned item representations lack full awareness of the session context. For example, in Figure~\ref{fig-app_verify}, although both sessions A and B have involved the flower, their intentions are quite different: one for decorating the garden while the other for decorating the kitchen. In contrast to blindly aggregating all neighbors' information to the flower without considering the session context~\cite{qiu2020exploiting, huang2021graph, zhang2021learning}, our approach learns session-aware item embeddings by selectively propagating information from relevant neighbors based on the current session. For instance, when determining which neighbors' information is aggregated to the flower, we recommend incorporating the lopper and watering can when decorating the garden in (b), while the sink and the table when decorating the kitchen in (c).

Secondly, the global transitional patterns among items in these hybrid models are typically constructed based on one type of interaction, such as co-purchase patterns~\cite{huang2021graph, qiu2020exploiting, pan2021graph, wang2020global}. However, on e-commerce platforms~\cite{xia2021knowledge}, items could form multiple relationships: substitution items are typically co-viewed and complementary items are typically co-add-to-carted (co-ATC)~\cite{wang2018path, mcauley2015inferring} by the same user. Uniformly using neighbor information may result in the dilution of diverse relationships among items and consequentially lead to unsatisfactory recommendations for users.

To overcome the above two challenges, we propose a knowledge graph-based SBRS with session-adaptive propagation. Our contributions are summarized as follows:
\begin{itemize}[leftmargin=*]
    \item For the first challenge, we propose a session-adaptive graph propagation to adaptively aggregate items' neighbor information based on the session contexts obtained by the transformer model.

    \item For the second challenge, we construct an item knowledge graph by extracting three different item co-relations with users. Additionally, we employ the heterogeneous graph transformer~\cite{hu2020heterogeneous} in the message-passing design to effectively aggregate neighborhood information based on the relationship type.

    \item We experimentally verify the proposed framework in enhancing existing SBRs, compare the performance improvement caused by different types of edges, and verify the session-adaptive propagation by visualizing the changing attentions of the same item in aggregating neighbors in different sessions.
\end{itemize}

    



\section{Related Work}\label{sec-relatedwork}
Earlier works in SBRSs leverage Markov Chains to infer the conditional probability of an item based on the previously interacted items~\cite{shani2005mdp, rendle2010factorizing}. 
More recent works have resorted to deep learning for session recommendation. Recurrent neural networks (RNNs)~\cite{xu2019recurrent, hidasi2018recurrent} such as GRU4Rec~\cite{hidasi2015session} and Transformers (TFs)~\cite{de2021transformers4rec} such as SASRec~\cite{kang2018self} have been developed to model the interactions among adjacent items in a session. 
To capture even more complex transitional patterns, graph-based methods such as SR-GNN~\cite{wu2019session} extract the session graph for each session and use a gated GNN to learn session embeddings. Different from previous works that only capture transitional patterns within the session, we construct a global graph with different types of edges to capture even more broad information, the related works of which are reviewed next.


Prior work explore global transitional patterns across different sessions by querying the global item graph~\cite{zhang2021learning, huang2021graph, qiu2020exploiting}. \cite{huang2021graph} designs a global context-enhanced inter-session relation encoder to capture the inter-session item-wise dependencies. \cite{zhang2021learning} constructs the dual session graph to model the pair-wise transition relationship between items based on the global connections. \cite{qiu2020exploiting} constructs the global graph by merging all individual session graphs. Recently, 
KSTT~\cite{zhang2021knowledge} uses 
an item-category knowledge graph for session recommendation. However, these models 
learn item embeddings from the global graph without 
session-tailored modification. Only GCE-GNN~\cite{wang2020global} and GCARM~\cite{pan2021graph} consider session adaptation in aggregating neighbors' information. However, GCE-GNN quantifies the importance of neighbors based on their similarity to the whole session without differentiating central items. GCARM treats all transitions similarly without distinguishing different interaction types. 
To handle these issues, we design a session-adaptive propagation to query neighbors based on session contexts and interaction types. The constructed KG in this work has only one type of node, the item, and we will consider other node types in the future. We put a more comprehensive review of related works in Appendix~\ref{app-related-work}.



\vspace{-1.25ex}
\section{The Proposed Framework}\label{sec-framework}
\begin{figure*}[t]
     \centering
     \hspace{-3ex}
     \includegraphics[width=1.01\textwidth]{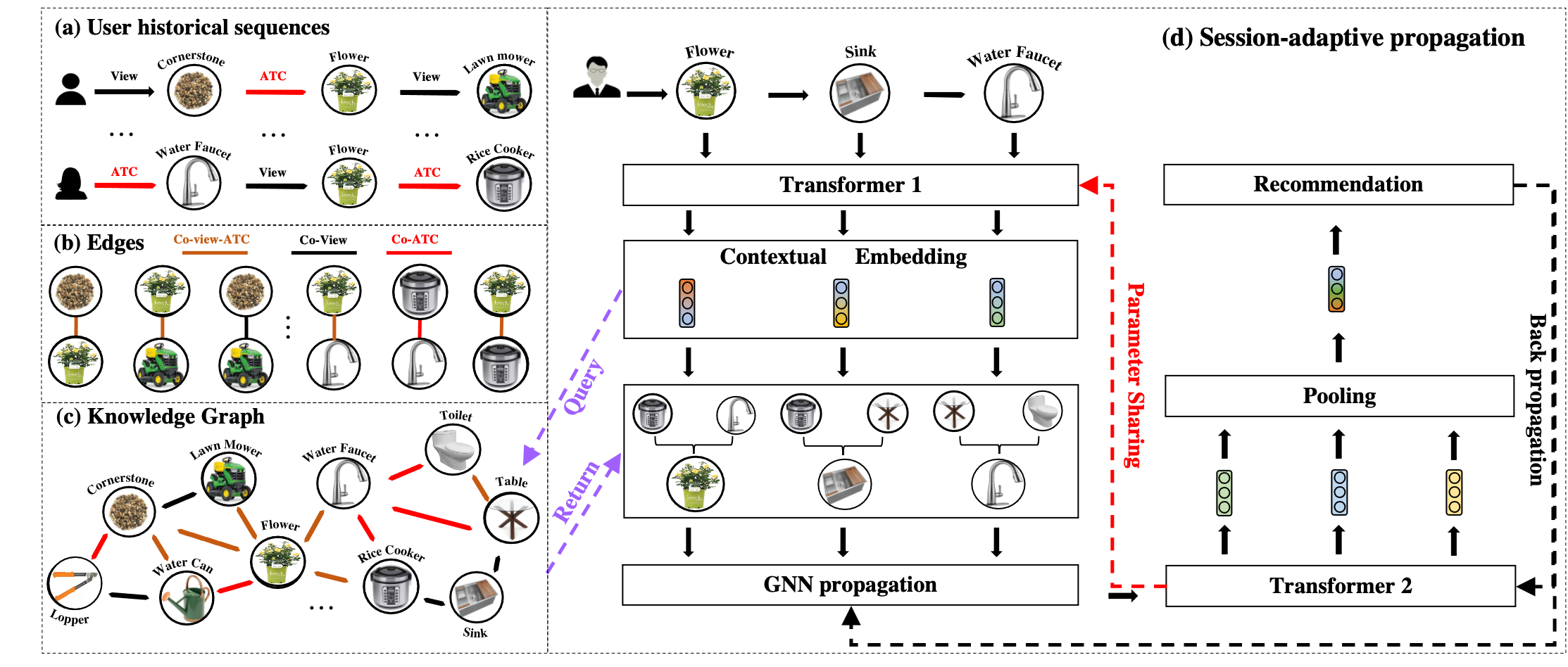}
     \vspace{-1.5ex}
     \caption{In (a)-(c), we first extract three types of edges from historical sessions to construct item knowledge graph. Then in (d), we forward the given session through the $1^{\text{st}}$ transformer layer to obtain items' contextual embeddings, which are used for query-relevant neighbors for GNNs to perform graph propagation. The propagated item embeddings are fed into the $2^{\text{nd}}$ transformer with a pooling layer afterward to obtain session embedding for the recommendation.}
     \label{fig-framework}
     \vspace{-1ex}
\end{figure*}

Our framework, as illustrated in Figure~\ref{fig-framework}, comprises of a GNN that obtains item embeddings by adaptively aggregating information from neighboring items based on the target session, and a transformer model that acquires session embeddings for predicting the next item. In the subsequent sections, we first explain the construction of the item knowledge graph, and then the GNN-based message passing model and the transformer-based prediction model.

\vspace{-1.25ex}
\subsection{Item Knowledge Graph Construction}\label{sec-graph}
As items typically exhibit two types of correlations, substitution and complementary \cite{zhang2021learning}, we extract three distinct types of edges, as depicted in Figure~\ref{fig-framework}(a)-(c), by examining whether two items co-occur within the same session: co-view, co-ATC, and co-view-ATC edges. For instance, in the first session shown in Figure~\ref{fig-framework}(a), the user first views the cornerstone, then adds the flower to the cart, and subsequently views the lawn mower. This sequence forms three edges: the co-view edge between the cornerstone and the lawn mower, the co-view-ATC edge between the cornerstone and the flower, and the co-view-ATC edge between the flower and the lawn mower. More formally, we define the edge weight from item $v_j$ to $v_i$ of type co-$t_1$-$t_2$ as follows: 
\vspace{-1ex}
\begin{equation}
\small
    w_{i \leftarrow j}^{(t_1, t_2)} = \frac{\sum_{m = 1}^{M}\mathbbm{1}(v_i, v_j\in \mathcal{S}_m, ~\tau(v_i, \mathcal{S}_m) = t_1, ~\tau(v_j, \mathcal{S}_m) = t_2)}{\sum_{m=1}^{M}\mathbbm{1}(v_i \in \mathcal{S}_m, ~\tau(v_i, \mathcal{S}_m) = t_1)},
\end{equation}
where $\mathcal{S}_m$ is the $m^{\text{th}}$-session, $M$ is the total number of sessions in the historical data, and $t_1, t_2\in\mathcal{T} = \{\text{view}, \text{ATC}\}$ and $\mathbbm{1}$ is an indicator function. Specifically, $\mathbbm{1}(v_i\in \mathcal{S}_m, v_j\in \mathcal{S}_m, \tau(v_i, \mathcal{S}_m) = t_1, \tau(v_j, \mathcal{S}_m) = t_2) = 1$ if \textbf{(a)} both $v_i, v_j$ belong to the $m^{\text{th}}$ session $\mathcal{S}_m$, \textbf{(b)} user interacts with $v_i$ following type $t_1$ and \textbf{(c)} interacts with $v_j$ following type $t_2$ in $\mathcal{S}_m$ and otherwise 0. Note that, we normalize edge weights based on the degree of the head node to avoid popular items from dominating the message-passing of GNNs~\cite{tang2020investigating, gupta2019niser}. 

Based on our empirical experiments, it was observed that the obtained graph may contain items with over a hundred neighbors. Therefore, in order to mitigate the heavy computational issue~\cite{ying2018graph} and prevent over-smoothing~\cite{chen2020simple} during the message passing, we sparsify the graph by retaining only the top-$K$ neighbors for each neighborhood type, based on the acquired edge weights. The statistics of the constructed networks are provided in Table~\ref{tab-data}.

\vspace{-4ex}
\subsection{Session-adaptive Propagation}\label{sec-GNN}
As mentioned earlier, a straightforward approach to utilize item graphs in SBRSs is to employ a GNN model to obtain item embeddings from the graph. These acquired embeddings can then be utilized in a transformer-based model to predict the next item based on the target session~\cite{zhang2021learning, huang2021graph, qiu2020exploiting}.
 
However, seen in Figure~\ref{fig-app_verify}, a key limitation of this architecture is that the contribution of each neighboring item remains unchanged in obtaining the center item's embedding, regardless of the session from which the item originates. As such, the item embeddings obtained from the GNN is unaware of the contextual information specific to the target session. This approach is sub-optimal since the purchasing intention associated with the same item naturally differs across different sessions, thereby necessitating changes in the selection of compatible neighbors based on the session-specific intention. Motivated by this observation, we propose a session-adaptive propagation that dynamically propagates neighbor information according to the item's unique context within each session.

To implement the above idea, given the target session, we input the initial item embedding within the session into a transformer, resulting in the generation of its session-aware representation. Subsequently, this obtained embedding is utilized to determine the edge weights for the GNN model. Finally, following the message-passing process, the updated item embeddings are once again fed into another transformer model to make the final prediction.

More formally, we first obtain item initial embeddings $\mathbf{E}^{1}$ by integrating item meta-attributes such as item title and category. Then, given an item $v_i$ in the session $\mathcal{S}_m$, we use the item initial embeddings $\mathbf{E}^{1}$ to obtain its contextual embedding $\mathbf{c}_i^m$:
\begin{equation}
    \mathbf{c}^{m}_i = transformer_{1}(v_i, \mathbf{E}^{1}, \mathcal{S}_m), \quad\forall v_i\in \mathcal{S}_m
\end{equation}
where $transformer_{1}$ is the 1$^{\text{st}}$ transformer. With the contextual embedding $\mathbf{c}_i^{m}$ as the query, inspired by~\cite{hu2020heterogeneous},  we perform heterogeneous graph transformer-based propagation to adaptively aggregate $v_i$'s neighbors' information relevant to the current session intention:
\begin{equation}\label{eq-heteroprop}
    \mathbf{h}_i^{m, l} = \frac{1}{|\mathcal{T}|^2}\sum_{t \in \mathcal{T}\times \mathcal{T}}||_{h=1}^{H}\sum_{v_j\in \mathcal{N}^{t}_i}\alpha_{i \leftarrow j}^{h, l, t, m}\mathbf{V}^{h, l, t}\mathbf{h}_j^{m, l - 1}
\end{equation}
\begin{equation}
\small
    {\alpha}^{h, l, t, m}_{i \leftarrow j} = \frac{(\mathbf{Q}^{h, l, t}\mathbf{h}_i^{m, l - 1})^{\top}(\mathbf{K}^{h, l, t}\mathbf{c}_j^{m})}{\sqrt{d/H}}, \forall v_i\in \mathcal{S}_m, \mathcal{S}_m\in\mathscr{S},
\end{equation}
where ${\alpha}^{h, l, t, m}_{i \leftarrow j}$ denotes the graph attention from item $v_j$ to $v_i$ under the head $h$, edge type $t$ at layer $l$. $\mathbf{Q}^{h, l, t}, \mathbf{K}^{h, l, t}, \mathbf{V}^{h, l, t}$ represent the query, key, and value matrix at the head $h$, edge type $t$, layer $l$ of graph attention. $H$ is the total number of heads. After $L$ layers graph transformer-based propagation, we obtain the final item embeddings $\mathbf{h}_i^{m, L}$. \textit{Since different session contexts provide different contextual embeddings $\mathbf{c}_i^m \ne \mathbf{c}_i^{m'}$, the calculated attention coefficients would also be different, i.e., $\alpha_{i \leftarrow j}^{h, l, t, m} \ne \alpha_{i \leftarrow j}^{h, l, t, m'}$. Then $\mathbf{h}_i^{m, L}$ and $\mathbf{h}_i^{m', L})$ would be different, and they would only include the neighborhood information that is relevant to the item's unique intention provided by the corresponding session context $\mathcal{S}_m (\mathcal{S}_{m'})$.}

To demonstrate the effectiveness of our proposed session-adaptive propagation in learning neighborhood attention based on the session context, we conduct a case study that visualizes the learned neighborhood attention for the same item across different sessions is indeed different. Further details are presented in Section~\ref{sec-casestudy}.

\subsection{Prediction and Optimization}\label{sec-bp}
After $L$ layers graph propagation, the item embeddings, e.g., $\mathbf{h}_i^{m, L}, \forall v_i \in \mathcal{S}_m$, are fed into $transformer_2$ with mean pooling to obtain the session embedding $\mathbf{S}_m$ and then we compute the cross-entropy loss to optimize the whole framework:
\begin{equation}\label{eq-predict}
\small
    \mathcal{L} = \sum_{m = 1}^{M}\sum_{i=1}^{|\mathcal{V}|}\mathbf{Y}_{m, i}\log\hat{\mathbf{Y}}_{m, i},~~~~\hspace{1ex} \mathbf{S}_m = transformer_2(\{\mathbf{h}_i^{m, L}|v_i \in \mathcal{S}_m\})
\end{equation}
where $\mathcal{V}$ is the total item space, $\mathbf{Y}_{m} \in \{0, 1\}^{|\mathcal{V}|}$ is the one-hot encoded next ground-truth item of the session $\mathcal{S}_m$, $\hat{\mathbf{Y}}_{m}=\sigma(p(\mathbf{S}_m))\in\mathbb{R}^{|\mathcal{V}|}$ is the predicted probability distribution of the next item for the session $\mathcal{S}_m$ from the linear prediction head $p$ followed by the softmax normalization $\sigma$. Note that $\mathbf{S}_m$ could be used in other tasks if paired with corresponding prediction heads, such as predicting the category/price of the next item as discussed in Section~\ref{sec-multitask}. The parameters of $transformer_{2}$ are optimized via back-propagation, and we periodically synchronize the parameters from $transformer_{1}$ to $transformer_{2}$~\cite{he2020momentum}. During the inference stage, we make the next item prediction by finding the item $i = \argmax_{j\in\mathcal{V}}\mathbf{Y}_{m, j}$.

\section{Experiments}\label{sec-experiment}
In this section, we verify the effectiveness of the proposed knowledge graph construction method and the session-adaptive propagation mechanism through extensive experiments. We first introduce the datasets and the experimental settings.

\subsection{Experimental Setup}
\subsubsection{Datasets}
We conduct experiments on the following datasets:
\begin{itemize}[leftmargin=*]
    \item \textbf{Diginetica}\footnote{https://competitions.codalab.org/competitions/11161} comes from CIKM Cup 2016. We follow the same pre-processing as~\cite{wu2019session}: sessions in the last and second-to-last week are used as testing and validation data. We filter out sessions of length less than 1 and items appearing less than 5 times. We extract item co-purchase edges from all sessions in \href{https://competitions.codalab.org/competitions/11161##learn_the_details-data2}{\textcolor{blue}{train-purchase.csv}} and item co-view edges from only training/validation sessions in \href{https://competitions.codalab.org/competitions/11161##learn_the_details-data2}{\textcolor{blue}{train-item-view.csv}} to avoid data leakage. The meta-attribute includes item title, category, and price.

    \item \textbf{Yoochoose 1/64}\footnote{https://www.kaggle.com/datasets/chadgostopp/recsys-challenge-2015} comes from the RecSys Challenge 2015. Sessions on the last and second last day are used as testing and validation data. We filter out sessions of length less than 1 and items appearing less than 5 times. We extract item co-purchase edges from all sessions in \href{https://www.kaggle.com/datasets/chadgostopp/recsys-challenge-2015}{\textcolor{blue}{yoochoose-buys.dat}} and item co-view edges from only training/validation sessions in \href{https://www.kaggle.com/datasets/chadgostopp/recsys-challenge-2015}{\textcolor{blue}{yoochoose-clicks.dat}} to avoid data leakage. Since items in this dataset have only one category attribute and it represents different meanings, e.g., 1-12 for item real categories, `S' for special offer, and 8-10 digit numbers for item brand, we only use item ID to demonstrate the effectiveness of session-adaptive heterogeneous propagation.

    \item \textbf{THD is a real industrial-level dataset} 
    from The Home Depot, the largest home improvement retailer in the USA. We sample 3,169,140/672,873/672,873 Add-to-Cart (ATC) sessions chronologically for constructing the train/valid/testing data. The multiplex graph is constructed by extracting co-view, co-ATC, and co-view-ATC edges. Items in this dataset have 7 meta-attributes: product title, hierarchical categories (i.e., L1, L2, L3, Leaf), brand, manufacturer, color, department, and class name.
\end{itemize}

Statistics of the three datasets are summarized in Table~\ref{tab-data}. Note that sparse sessions refer to the ones containing items appearing less than 5 times in the whole dataset. These sessions are used to evaluate our framework on sessions including cold-start items.

\begin{table}[t!]
\scriptsize
\setlength{\extrarowheight}{.08pt}
\setlength\tabcolsep{0.3pt}
\caption{Statistics of datasets used for experiments and their corresponding knowledge graphs.}
\vspace{-2ex}
\centering
\label{tab-data}
\begin{tabular}{lccccc}
\hline
\textbf{Dataset} & \textbf{Train/Val/Test Seqs} & \# \textbf{Edges} & \# \textbf{Nodes} & \textbf{Sparse Seqs} & \textbf{Meta data} \\
\hline
Diginetica & \makecell{675,673/\\43,541/\\68,571} & 1,576,571 & 123,273& 13,867 & Title/Category/Price \\
\hline
\makecell[l]{Yoochoose \\ 1/64} & \makecell{369,142/\\45,864/\\55,898} & 1,025,176 & 52,739 & 15,385 & -- \\
\hline
\makecell[l]{The Home \\ Depot (THD)} & \makecell{3,169,140/\\672,873/\\672,873} & 4,162,712 & 1,317,149 & 227,941 & \makecell{Title/Category/Brand/\\Color/Manufacturer/Class/\\Department}\\
\hline
\end{tabular}
\vspace{-2ex}
\end{table}

\begin{table*}[t]
\small
\caption{Performance comparison (\%) of utilizing meta-attribute embedding layer and session-adaptive heterogeneous propagation layer. The best and runner-up are in \textbf{bold} and \underline{underlined}. Note that N(M)@10 represents NDCG(MRR)@10.}
\vspace{-1ex}
\setlength{\extrarowheight}{.05pt}
\setlength\tabcolsep{4pt}
\label{tab-performance}
\begin{tabular}{ll|cccc|cccc|cccc|c}
\hline
\multicolumn{2}{c|}{\multirow{3}{*}{\textbf{Backbone}}} & \multicolumn{4}{c|}{\textbf{Diginetica}} & \multicolumn{4}{c|}{\textbf{Yoochoose 1/64}} & \multicolumn{4}{c|}{\textbf{THD}} & \multirow{3}{*}{\textbf{Increase} ($\uparrow$)} \\
\multicolumn{2}{c|}{} & \multicolumn{2}{c}{N@10} & \multicolumn{2}{c|}{M@10} & \multicolumn{2}{c}{N@10} & \multicolumn{2}{c|}{M@10} & \multicolumn{2}{c}{N@10} & \multicolumn{2}{c|}{M@10} &  \\
 &  & Full & Sparse & Full & Sparse & Full & Sparse & Full & Sparse & Full & Sparse & Full & Sparse &  \\
\hline
\multirow{3}{*}{\textbf{RNN}} & GRU4Rec & 12.66 & 11.04 & 9.55 & 8.19 & 32.77 & 28.32 & 26.52 & 22.81 & 14.40 & 7.66 & 12.34 & 6.53 & -- \\
 & GRU4Rec$_m$ & 12.38 & 10.62 & 9.27 & 7.77 & -- & -- & -- & -- & 17.09 & 10.42 & 14.19 & 8.41 & 10.55$\%$ \\
 & GRU4Rec$_m*$ & 13.01 & 11.47 & 9.73 & 8.37 & -- & -- & -- & -- & 18.13 & 11.33 & 15.23 & 9.34 & 18.88$\%$ \\
 \hline
\multirow{3}{*}{\shortstack{\textbf{Trans-}\\\textbf{former}}} & SASRec & 14.70 & 13.00 & 11.11 & 9.57 & 33.68 & 28.59 & 27.16 & 22.87 & 15.27 & 8.11 & 12.97 & 6.83 & -- \\
 & SASRec$_m$ & 14.68 & 12.79 & 11.08 & 9.28 & -- & -- & -- & -- & 17.87 & 11.11 & 14.87 & 9.01 & 11.94$\%$ \\
 & SASRec$_m*$ & 15.31 & 13.68 & 11.60 & 10.12 &  &  & -- & -- & 18.53 & 11.72 & 15.53 & 9.64 & 18.28$\%$ \\
 \hline
\multirow{5}{*}{\textbf{Graph}} & KGHT & 17.73 & 15.99 & 13.33 & 11.74 & \underline{34.81} & \underline{30.14} & \underline{28.04} & \underline{24.10} & 16.59 & 9.29 & 14.02 & 7.74 & -- \\
 & KGHT$_m$ & 17.70 & 15.78 & 13.27 & 11.52 & -- & -- & -- & -- & 18.47 & 11.68 & 15.49 & 9.59 & 8.45\% \\
 & KGHT$_m*$ & 17.81 & \underline{16.88} & 13.38 & \underline{12.49} & -- & -- & -- & -- & \underline{18.78} & \underline{11.89} & \textbf{15.74} & \underline{9.79} & 11.59$\%$ \\
 & KGHT$_q$ & \underline{18.64} & 16.74 & \underline{13.97} & 12.28 & \textbf{35.80} & \textbf{31.71} & \textbf{28.80} & \textbf{25.21} & 17.30 & 10.20 & 14.45 & 8.35 & 4.25$\%$ \\
 & KGHT$_{qm*}$ & \textbf{18.93} & \textbf{17.16} & \textbf{14.22} & \textbf{12.65} & -- & -- & -- & -- & \textbf{18.83} & \textbf{12.17} & \underline{15.67} & \textbf{9.91} & 14.10$\%$ \\
 \hline
\end{tabular}

\begin{tablenotes}
    \centering 
      \footnotesize
      \item \textbf{X}: Backbone X using item ID but no meta-attributes;\quad \textbf{X}$_{m}$: Backbone X using item meta-attributes but no ID; \textbf{X}$_{m*}$: Backbone X using both item meta-attributes and ID;
      \item \textbf{X}$_{q}$: Backbone X using Graph and Session-adaptive propagation; \textbf{X}$_{qm*}$: Backbone X using Graph, Session-adaptive propagation, and meta-attribute.
\end{tablenotes}
\vspace{-1ex}
\end{table*}

\subsubsection{Item embedding initialization} 
We initialize item embeddings based on their meta-attributes. We use different embedding techniques to extract different types of item metadata~\cite{shalaby2022m2trec}, which can be numerical, categorical, and textual. For each numerical attribute, we directly embed it as a real-valued number. For each categorical attribute, we initialize a unique learnable embedding matrix. For textual attributes such as title and description, we first construct the token embedding matrix and then the title/description embedding is computed by mean pooling over the embeddings of corresponding tokens in that sentence. Note that pre-trained NLP models are not preferred here to avoid capturing noisy semantic signals. For example, even though silver sinks and creamy white stones share semantic-similar colors, they are essentially purposed for decorating different rooms. We empirically observe that utilizing this token-based embedding avoids capturing noisy semantic signals since they correspond to different tokens. We concatenate different types of embeddings to form the final item meta-embeddings and feed them into the transformer model.

\subsubsection{Backbones} Note that our constructed knowledge graph and the proposed session-adaptive propagation can be applied to enhance any embedding-based SBRS. To demonstrate this, we select three fundamentally different but representative backbones and equip them with our framework:

\begin{itemize}[leftmargin=*]
    \item \textbf{GRU4Rec}~\cite{hidasi2015session}: The very first model leveraging RNNs to characterize item sessions for session recommendation.

    \item \textbf{SASRec}~\cite{kang2018self}: The very first model leveraging the self-attention from the transformer to draw context from all user-item interactions in the same session.

    \item \textbf{KGHT}~\cite{xia2021knowledge}: A graph-based model constructing the item relational graph and leveraging graph attention to capture item relations for recommendations. Since this work is not initially designed for session recommendation, we modify it to align with our problem setting.
\end{itemize}

Equipping each of the above three backbones with the item meta-attribute embedding layer and the session-adaptive propagation layer, we end up with 11 model configurations. We name each new configuration by combining the name of its backbone
and the equipped techniques, e.g., GRU4Rec$_{m}$ denotes the backbone of GRU4Rec with item meta-attribute embedding layer, SASRec$_{m*}$ denotes the backbone of SASRec with both item meta-attribute and ID, and KGHT$_{q}$ denotes the backbone of KGHT equipped with session-adaptive propagation layer. The architecture of each model configuration is summarized in Table~\ref{tab-architecture}.





\vspace{-1ex}
\subsubsection{Evaluation Metric and Implementation Details} Following~\cite{wu2020sse, jin2023dual}, we report the average of NDCG@10 (N@10) and MRR@10 (M@10) over all sequences in the test set. We assign a dedicated embedding layer for each attribute, and the embedding dimension is determined based on the total number of distinct tokens of the corresponding attribute vocabulary. To ensure a fair comparison, we tune the following hyperparameters for each model individually: the dimension of ID embedding layer $\{64, 128, 256\}$, the number of layers for self-attention and graph propagation $\{1, 2, 3\}$, the number of attention heads $\{1, 4, 8\}$, the dimension of hidden embeddings $\{100, 512\}$, learning rate $\{1e^{-4}, 1e^{-3}, 1e^{-2}\}$, the L2 penalty $\{0, 5e^{-4}\}$, training epochs $\{100, 200\}$, the batch size $\{100, 512\}$, dropout ratio $\{0.1, 0.25\}$ across all models. We save the model performing best on validation sessions and evaluate it on testing sessions. Training epochs for Diginetica/Yoochoose/THD are set to be 200/200/30. We synchronize the parameters of $transformer_2$ from $transformer_1$ every epoch. 

\begin{table}[t!]
\small
\setlength{\extrarowheight}{.095pt}
\setlength\tabcolsep{2.25pt}
\centering
\caption{An architecture comparison of different backbones.}\label{tab-architecture}
\vspace{-2ex}
\begin{tabular}{lccccc}
\hline
\textbf{Backbone} & \textbf{ID}& \textbf{\makecell{Meta- \\attribute}} & \textbf{Graph} & \textbf{\makecell{Session-adaptive \\propagation}} & \textbf{\makecell{Multi-task \\learning}}\\
 \hline
\textbf{GRU4Rec} & \cmark & \xmark & \xmark & \xmark & \xmark\\  
\textbf{GRU4Rec}$_m$ & \xmark & \cmark & \xmark & \xmark  & \xmark\\
\textbf{GRU4Rec}$_m*$ & \cmark & \cmark & \xmark & \xmark & \xmark\\
\textbf{M-GRU4Rec}$_m*$ & \cmark & \cmark & \xmark & \xmark & \cmark\\
\hline
\textbf{SASRec} & \cmark & \xmark & \xmark & \xmark & \xmark\\
\textbf{SASRec}$_m$ & \xmark & \cmark & \xmark & \xmark  & \xmark\\
\textbf{SASRec}$_m*$ & \cmark & \cmark & \xmark & \xmark & \xmark\\
\textbf{M-SASRec}$_m*$ & \cmark & \cmark & \xmark & \xmark & \cmark\\
\hline
\textbf{KGHT} & \cmark & \xmark & \cmark & \xmark & \xmark\\
\textbf{KGHT}$_m$ & \xmark & \cmark & \cmark & \xmark & \xmark\\
\textbf{KGHT}$_m*$ & \cmark & \cmark & \cmark & \xmark  & \xmark\\
\textbf{KGHT}$_q$ & \cmark & \xmark & \cmark & \cmark  & \xmark\\
\textbf{KGHT}$_{qm*}$ & \cmark & \cmark & \cmark & \cmark  & \xmark\\
\textbf{M-KGHT}$_{qm*}$ & \cmark & \cmark & \cmark & \cmark  & \cmark\\
 \hline
\end{tabular}
\vspace{-3.5ex}
\end{table}

\subsection{Model Configuration Analysis}\label{sec-ablation}
To demonstrate the effectiveness of the proposed knowledge graph and session-adaptive propagation, we compare three backbones GRU4Rec, SASRec, and KGHT with their corresponding enhanced versions in Table~\ref{tab-performance}. For brevity, we represent any of the three backbones as $X$ in the following text:
\begin{itemize}[leftmargin=*]
    \item Compared with $X$, $X_{m*}$ incorporates item meta-attributes and improves the performance by around $12\%-19\%$ on average. This is because ID-based embedding only captures the topological proximity of each item to all other items and cannot provide generalizability, especially when items' topological information is noisy/sparse. Leveraging meta-attributes alleviates this issue by transferring the learned information among items sharing the same meta-attribute. Since THD has more abundant types of well-curated meta-attributes than the ones of Diginetica, as evidenced in Table~\ref{tab-data}, $X_{m*}$ achieves an even larger performance gain over $X$ on THD than on Diginetica.\vspace{1ex}

    \item Compared with $X$, $X_{m}$ achieves comparable and sometimes slightly worse performance on Diginetica, e.g., $9.55\%$ for GRU4Rec while $9.27\%$ for GRU4Rec$_m$ in MRR@10. This is because solely relying on meta-attributes to represent items may lose topological information in the sessions. Two items sharing the same meta-attributes will be encoded the same, even though they may be involved in significantly different sessions. This is also evidenced by the better performance of $X_m{*}$ than $X_m$ after combining both item ID and item meta-attribute. Different from Diginetica, $X_m$ always achieves higher performance than $X$ on THD because more abundant meta-attributes there enable the concatenated embeddings to be more unique and hence can somewhat mimic the function of Item ID in capturing topological information embedded in the sessions.\vspace{1ex}

    \item Comparing among different backbones, graph-based models achieve higher performance than non-graph-based ones, which aligns with the notion that incorporating global information across different sessions is conducive to the recommendation~\cite{zhang2021learning, huang2021graph, qiu2020exploiting}. More specifically, KGHT$_q$ gains 4.25$\%$ improvement over KGHT because the designed session-adaptive propagation only aggregates the most relevant neighbor information to the session context and avoids introducing unrelated neighbors. 
    
    
\end{itemize}

\subsection{Influence of Different Types of Edges}\label{sec-edge}

We further analyze the influence of co-view, co-ATC and co-view-ATC edges in the constructed knowledge graph. Specifically, we use \textbf{KGHT}$_{qm*}$ as the baseline and respectively remove three types of edges, e.g., \textbf{KGHT}$_{qm*}$ w/o v removes the co-view edges and \textbf{KGHT}$_{qm*}$ w/o diff treat different types of edges uniformly by employing the same graph attention layer\footnote{Instead of averaging aggregated embeddings across all $\mathcal{T}^2$ types of edges in Eq~\eqref{eq-heteroprop}, we use a uniformed query, value and key matrices.}. The performance of removing each specific type of edges is reported in Table~\ref{tab-edge} and we draw three observations:

\begin{table}[t]
\small
\caption{Ablation study on different types of edges.}
\setlength{\extrarowheight}{.05pt}
\setlength\tabcolsep{2pt}
\label{tab-edge}
\centering
\vspace{-1ex}
\begin{center}
\begin{tabular}{l|cc|cc|cc|c}
\hline
\multirow{2}{*}{\textbf{Backbone}} & \multicolumn{2}{c|}{\textbf{Diginetica}} & \multicolumn{2}{c|}{\textbf{Yoochoose 1/64}}& \multicolumn{2}{c|}{\textbf{THD}} & \multirow{2}{*}{\textbf{Decrease} ($\downarrow$)}\\
& N@10 & M@10 & N@10 & M@10 & N@10 & M@10 & \\
\hline
\textbf{KGHT}$_{qm*}$ & 18.94 & 14.22 & 35.80 & 28.80 & 18.83 & 15.67 & --- \\
\hspace{2ex}-- w/o diff & 18.34 & 13.80 & 35.15 & 28.22 & 18.68 & 15.50 & 2.02$\%$\\
\hspace{2ex}-- w/o v & 15.21 & 11.46& 34.36 & 27.68 & 18.66 & 15.53 & 9.78$\%$\\
\hspace{2ex}-- w/o a & 18.70 & 13.97 & 35.04 & 28.09 & 18.76 & 15.64 & 1.39$\%$\\
\hspace{2ex}-- w/o va & / & / & / & / & 18.79 & 15.63 & 0.23$\%$\\
 \hline
\end{tabular}
\end{center}
\begin{tablenotes}
    \centering
      \footnotesize
      \item \textbf{*} \textbf{w/o diff}: uniformly aggregate different types of neighbors with no differentiation. 
      \item \textbf{*} \textbf{w/o v/a/va}: with no co-view/co-ATC/co-view-ATC edges.
\end{tablenotes}
\end{table}

\begin{itemize}[leftmargin=*]
    \item On Diginetica and Yoochoose 1/64, removing co-view edges decreases the performance most because of the following two reasons. First, sessions in these two datasets track user click activities rather than purchase activities and consist of more substitution items rather than complementary ones~\cite{wang2018path, mcauley2015inferring}. Therefore, removing co-view edges that essentially capture the substitution relationship between any two items hurts the performance more than removing other types of edges. Secondly, co-view edges are constructed from both training and validation sessions, message-passing along which captures more recent transitional patterns encoded in the validation sequences. The performance decrease here indicates the distribution shift in transitional patterns from training sessions to validation sessions. One promising direction is to treat sessions at different time stamps differently, as recent transition patterns may be more indicative of the future sessions than the past ones~\cite{hu2020open}.

    \item Conversely, on THD dataset, since the session is composed of the user's sequentially add-to-cart activity, removing co-view edges causes minor performance degradation compared with the one on the other two datasets. Moreover, removing co-view-ATC edges causes little-to-no performance change, which indicates that capturing the view-to-ATC transition patterns may not help predict users' next clicked items.

\end{itemize}

\subsection{Multi-task Learning}\label{sec-multitask}
In this section, we explore the impact of squeezing item meta-label information such as taxonomy and price into the item embeddings on the performance of next item prediction. Concretely, we feed the session embedding $\mathbf{S}_m$ obtained from Eq.~\eqref{eq-predict} into task-specific prediction heads to predict the next item meta-labels. The loss of predicting next-item meta-labels is combined with the loss of next-item prediction to jointly train the whole model and hence can be essentially deemed as self-supervised learning~\cite{jin2020self, hgmae}. We first present the performance of next-item prediction before/after incorporating multi-task learning in Table~\ref{tab-multinextitem}. Specifically. \textbf{M-X} denotes the basic model \textbf{X} augmented by the multi-task learning. The item meta-labels used in Diginetica and THD are the price and L1 categories, respectively.

As shown in Table~\ref{tab-multinextitem}, in most cases, jointly training models by predicting the next-item meta-labels improves next-item prediction. This indicates that squeezing knowledge of the next-item meta-label could sometimes help next-item prediction. An exceptional case is applying \textbf{M-KGHT}$_{qm*}$ to Diginetica; the performance does not always increase because aggregating the non-informative meta-attributes of items by message-passing in Diginetica causes little-to-no benefit in next-item prediction. The performance improvement on THD achieved by multi-task learning is stronger than the one on Diginetica because meta-labels of items at THD are more carefully annotated and hence can uniquely characterize the functionality of corresponding items.

\begin{table}[t]
\small
\caption{Performance improvement by multi-task learning.}
\setlength{\extrarowheight}{.05pt}
\setlength\tabcolsep{2pt}
\label{tab-multinextitem}
\vspace{-0.75ex}
\centering
\begin{tabular}{l|cc|cc}
\hline
\multirow{3}{*}{\textbf{Backbone}} & \multicolumn{2}{c|}{\textbf{Diginetica}} & \multicolumn{2}{c}{\textbf{THD}} \\
 & N@10 & M@10 & N@10 & M@10 \\
 & (All, Sparse) & (All, Sparse) & (All, Sparse) & (All, Sparse) \\
 \hline
 \textbf{GRU4Rec}$_{m*}$ & (13.01, 11.47) & (9.73, 8.37) & (18.13, 11.33) & (15.23, 9.34)\\
\textbf{M-GRU4Rec}$_{m*}$ & (13.05, 11.57) & (9.74, 8.48) & (18.61, 11.76) & (15.66, 9.71)\\
\hline
\textbf{SASRec}$_{m*}$ & (15.31, 13.68) & (11.60, 10.12) & (18.54, 11.72) & (15.53, 9.64)\\
\textbf{M-SASRec}$_{m*}$ & (15.66, 13.84) & (11.83, 10.18) & (18.80, 12.08) & (15.78, 9.97)\\
\hline
\textbf{KGHT}$_{qm*}$ & (18.93, 17.16) & (14.22, 12.65) & (18.83, 12.17) & (15.67, 9.91) \\
\textbf{M-KGHT}$_{qm*}$ & (18.95, 17.11) & (14.23, 12.53) & (18.98, 12.28) & (15.81, 10.02)\\
\hline
\end{tabular}
\end{table}

\begin{figure*}[htbp!]
     \centering
     \includegraphics[width=1\textwidth]{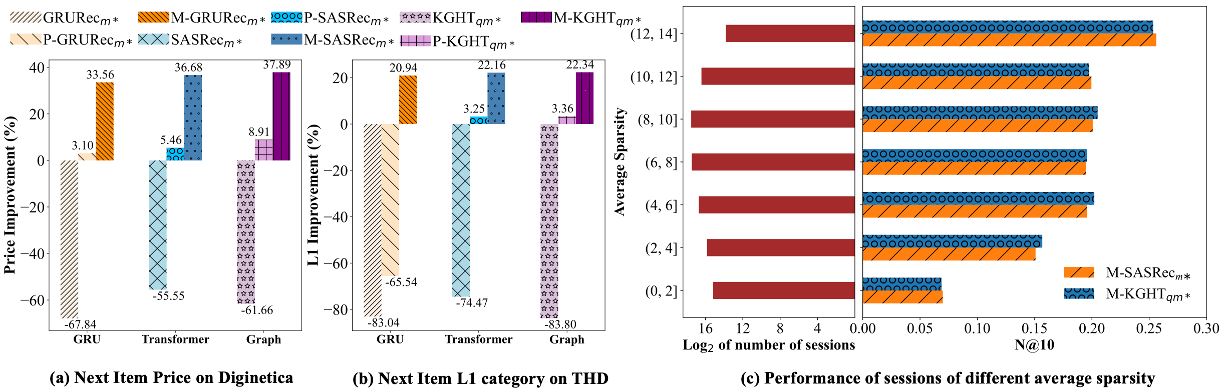}
     \vspace{-3ex}
     \caption{(a)-(b): Comparing the performance improvement(\%) over the baseline Top-N on predicting next-item meta-labels. \textbf{Top-N}: the simple heuristic recommending the most frequent meta-labels among previous items within one session; (2) \textbf{X$_{m*}$} uses an untrained prediction head to predict the task-specific label of the next item; (3) \textbf{P-X$_{m*}$} first predicts top-N next items and use their task-specific labels as recommendations; (4) \textbf{M-X$_{m*}$} uses a trained prediction head that is trained by multi-task learning to predict the task-specific label of the next item. (c) N@10 on sessions of different sparsity. M-KGHT$_{qm*}$ is better than M-SASRec$_{m*}$ on sessions of the sparsity at the middle level.}
     \label{fig-multi-task}
     \vspace{-2ex}
\end{figure*}

Predicting next-item meta-labels by itself has many real-world applications, such as previewing the users' purchasing budget if predicting the next-item price~\cite{lashinin2022next} and summarizing the users' purchasing intention if predicting the next-item category~\cite{cai2021category}. Therefore, we also compare the performance of next-item meta-label prediction with some other baselines, including \textbf{Top-N}, \textbf{X$_{m*}$}, \textbf{P-X$_{m*}$}, and \textbf{M-X$_{m*}$}. Their detailed definitions are included in the caption of Figure~\ref{fig-multi-task}(a)-(b). We report the performance improvement of each model over the baseline \textbf{Top-N}. As shown in Figure~\ref{fig-multi-task}(a)-(b), training prediction heads by multi-task learning significantly enhances the next-item meta-label prediction. The performance improvement of N@5 for predicting price on Diginetica of GRU4Rec increases from -67.84\% to +33.56\%, and N@5 for L1 category prediction on THD of SASRec increases from -74.47\% to +22.16\%. Compared among different backbones, graph-based models are better than transformer-based ones and further than GRU-based ones, showing that capturing global transitional patterns is also conducive to predicting the next item meta-label.

\subsection{Analysis on Sessions of Different Sparsity}\label{sec-length}
Prior research has identified degree-related biases in various contexts, including node classification~\cite{tang2020investigating}, link prediction~\cite{li2021user}, knowledge graph completion~\cite{shomer2023toward}, and conventional non-session-based recommender systems~\cite{wang2022collaboration, wang2022degree}. Analogically, we assess the relationship between the next-item prediction performance of the session and its sparsity, defined as the average degree of items in that session. In essence, higher average sparsity in a session indicates lower average involvement of items from that session in other sessions and a reduced number of global transitional patterns that the global knowledge graph can capture. We group sessions based on their average sparsity and report the average N@10 of M-SASRec$_m*$ and M-KGHT$_{qm*}$ across all sessions for each group in Figure~\ref{fig-multi-task}(c). 

Clearly, M-KGHT$_{qm*}$ performs better than M-SASRec$_{m*}$ when average sparsity ranges between 2-10 while worse when average sparsity is either below 2 or above 10. On one hand, items in sessions with extremely low sparsity participate in fewer sessions and thus, the captured global transitional patterns across different sessions are insufficient. Therefore, relying more on neighbors as M-KGHT$_{qm*}$ brings no advantages compared with relying more on meta-attributes as M-SASRec$_{m*}$. On the other hand, items in sessions with extremely higher sparsity generally have higher degrees and are usually co-view/co-ATC with many different types of items. Aggregating information from various types of neighboring items serving different functionalities may corrupt the intention of the current session. Because the number of sessions of sparsity between [2, 10] is higher than the number of other sessions, the overall performance of M-KGHT$_{qm*}$ is still higher than M-SASRec$_{m*}$. Further work can study this phenomenon from the fairness perspective and diversity perspective~\cite{dong2023fairness, zhaok2021autoemb, liu2020automated} and focus on balancing the performance of sessions of various sparsity by designing adaptive embedding sizes for items in different sessions~\cite{zhaok2021autoemb, liu2020automated} based on their involvement in different sessions. In addition, we can define session diversity as the average involvement of items and adapt existing diversity methods into alleviate session unfairness.

\begin{figure*}[t!]
     \centering
     \hspace{-3ex}\includegraphics[width=0.95\textwidth]{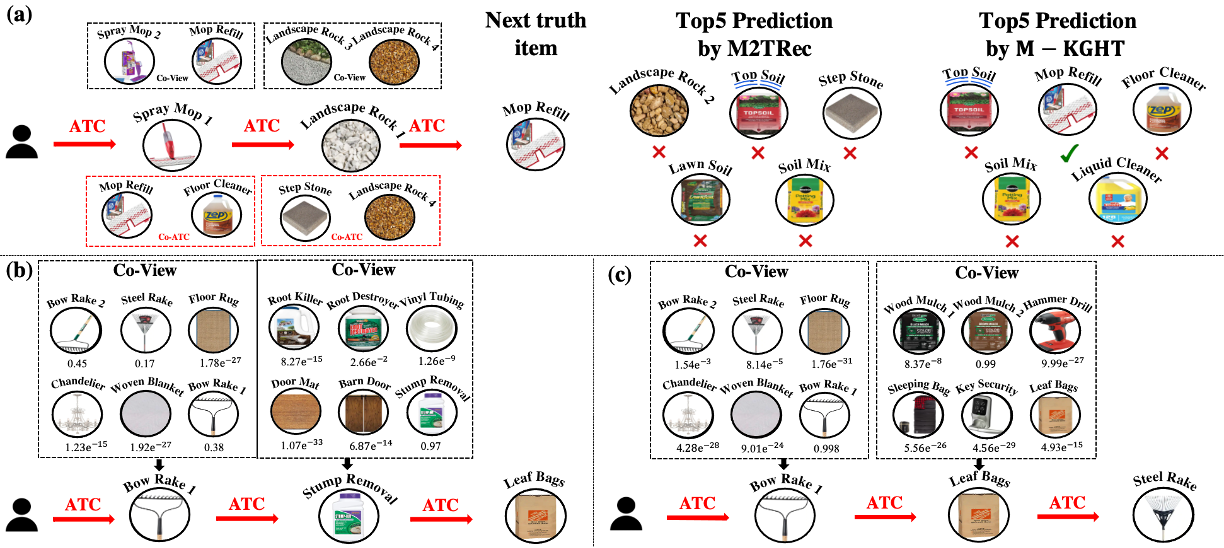}
     \caption{Case study of our proposed \textbf{M-KGHT}. (a) Comparing the Top5 recommendation by M2TRec and \textbf{M-KGHT}. By leveraging neighborhood information of Spray Mop1, the correct item Mop Refill appears in the recommendation list. (b)-(c) visualizes the learned attention of one attention head over co-view neighbors. Since users in both of these two sessions intend to clean the gardens, the session-adaptive graph propagation successfully learns the higher intention for garden-related items.}
     \vspace{-1.75ex}
     \label{fig-example}
\end{figure*}

\begin{table}[t!]
\small
\caption{Performance comparison among different baselines.}
\vspace{-2ex}
\centering
\label{tab-perform}
\begin{tabular}{l|cc|cc|cc}
\hline
\multirow{2}{*}{\textbf{Baseline}} & \multicolumn{2}{c|}{\textbf{Diginetica}} & \multicolumn{2}{c|}{\textbf{Yoochoose 1/64}} & \multicolumn{2}{c}{\textbf{THD}} \\
 & N@10 & M@10 & N@10 & M@10 & N@10 & M@10  \\
\hline
\textbf{NARM} & 16.06 & 12.12 & 34.45 & 28.10 & 17.44 & 14.62\\
\textbf{SR-GNN} & \underline{17.10} & \underline{12.82} & \underline{35.44} & \underline{28.66} & 17.08 & 14.45\\
\textbf{M2TRec} & 15.66 & 11.83 & 33.68 & 27.16 & \underline{18.80} & \underline{15.78}  \\
\hline
\textbf{M-KGHT}$_{qm*}$ & \textbf{18.95} & \textbf{14.23} & \textbf{35.80} & \textbf{28.80} & \textbf{18.98} & \textbf{15.81} \\
\hline
\end{tabular}
\vspace{-3ex}
\end{table}

\subsection{Performance Comparison with baselines}
From the analysis in previous sections, we select the best model configuration: M-KGHT$_{qm*}$ and compare it with state-of-the-art baselines \textbf{NARM}~\cite{li2017neural}, \textbf{SR-GNN}~\cite{wu2019session} and \textbf{M2TRec}~\cite{shalaby2022m2trec}. For the implementation of M2TRec and M-KGHT$_{qm*}$ on Yoochoose 1/64 with no item meta-attributes, we directly use item ID as the input. We modify the implementation of \href{https://github.com/Wang-Shuo/Neural-Attentive-Session-Based-Recommendation-PyTorch}{NARM} and \href{https://github.com/CRIPAC-DIG/SR-GNN}{SR-GNN} to include the validation performance and hence align with our experimental setting. In Table~\ref{tab-perform}, M2TRec performs worse than SR-GNN and NARM on Diginetica and Yoochoose 1/64 while better on THD. This is because M2TRec designs an item meta-embedding layer to integer item meta-features and the more informative meta-features on THD than the ones on Diginetica/Yoochoose maximize the benefit of using M2TRec. However, our proposed M-KGHT$_{qm*}$ achieves the best performance across all three datasets for both NDCG@10 and MRR@10. This exhibits the general ability of the proposed framework to realize superior performance over existing methods across datasets with varying real-world dynamics, i.e., having varying amounts of meta-attributes, user-item interaction types, and session lengths. Note that because THD is an industrial-scale dataset (THD has around 3 million training sessions that are 5/10 times larger than Diginetica/Yoochoose), the performance gain on THD is slightly weaker than the gains on the other two datasets.

\subsection{Industrial-level Case Study}\label{sec-casestudy}
We further deploy our designed system in the industrial-level setting by training it with 6 million sessions spanning the last two years and evaluating over 0.1 million sessions in the following month. We achieved 2\% performance improvement over the previously deployed model at The Home Depot. Next, we conduct some case studies to visualize the effect of our proposed system.


\vspace{-0.5ex}
\subsubsection{Visualizing the recommendation of M2TRec and M-KGHT$_{qm*}$}
To interpret the advantage of the proposed \textbf{M-KGHT}$_{qm*}$ over the second best model \textbf{M2TRec} on THD dataset, we select the sessions where the top-5 recommendation list given by \textbf{M-KGHT}$_{qm*}$ hit the next truth item while the one given by \textbf{M2TRec} does not. We visualize one example in Figure~\ref{fig-example}(a) where the customer sequentially add-to-cart the spray mop and landscape rock. Items on the top-5 recommendation list given by \textbf{M2TRec} are uniformly aligned with the landscape rock while recommended items by \textbf{M-KGHT}$_{qm*}$ align with both the spray mop and landscape. Furthermore, because \textbf{M-KGHT}$_{qm*}$ aggregates neighbor information of mop refill to spray mop, the recommendation hits the true item, which demonstrates the benefits of leveraging neighborhood information.

\vspace{-1ex}
\subsubsection{Visualizing the attention of session-adaptive propagation}
We further visualize the graph attention learned by our session-adaptive propagation in Figure~\ref{fig-example}(b)-(c). Clearly, users generating these two sequences (b)-(c) intend to clean their own gardens and successfully, the model learns to aggregate less information from irrelevant neighbors, e.g., $1.78e^{-27}$ from floor rug to bow rake 1 in (b) and $9.99e^{-27}$ from hammer drill to leaf bags in (c). Interestingly, we find model sometimes pays attention to only one relevant neighbor. For example, even though both bow rake 1 and steel rake are aligned with the intention of our second customer in (c), the model pays its whole attention to bow rake 1. This aligns with diversity observation in~\cite{vaswani2017attention} and motivates the design of multi-head attention to focus on different important neighbors. More importantly, we can find the neighborhood attention of the same item bow rake 1 varies from 0.38 in (b) to 0.998 in (c), demonstrating that even though the neighborhoods are the same, the attention assigned to them changes if the sequence content changes. This verify the effectiveness of our proposed session-adaptive propagation.

\vspace{-2ex}
\section{Conclusion}\label{sec-conclusion}
Characterizing user intention by modeling global transitional patterns of user-item interactions is essential in the session-based recommendation. Traditional transformer-based models fail to capture global transitional patterns among items. More recent GNN-augmented transformers ignore the session context and only consider one type of customer-item interaction. Given these problems, we propose a knowledge graph-based session recommendation framework with session-adaptive propagation. We construct the graph by extracting three different types of user-item interactions and design a session-adaptive propagation for aggregating neighbors' information based on their consistency with the session intention. A comprehensive ablation analysis shows the proposed strategies provide a 10\%-20\% improvement. Moreover, our case study on recommendation interpretation demonstrates that learned neighborhood attention is highly determined by the consistency of the neighbor with the session intention. 


\section*{Acknowledgements} This research is supported by The Home Depot and the National Science Foundation (NSF) under grant number IIS2239881.

\bibliographystyle{ACM-Reference-Format}
\bibliography{references}

\appendix
\newpage 
\section{Appendix}
\subsection{Comprehensive Related Work}\label{app-related-work}
\subsubsection{Session-based Recommendation System}
Given that not all historical user-item interactions are beneficial to predict users' current preferences~\cite{wang2021survey}, SBR Systems have emerged with increasing attention in recent years~\cite{hidasi2015session, hidasi2018recurrent}. Earlier works leverage Markov Chains to infer the conditional probability of an item based on the previously interacted items~\cite{shani2005mdp, rendle2010factorizing}. \cite{shani2005mdp} infers the transition function based on user data and additionally employs skipping/clustering to enhance the recommendation. \cite{rendle2010factorizing} factorizes the probability transition matrix of each user to model sequential behavior between every two adjacent clicks. More recent works have resorted to deep learning for session recommendation. Recurrent neural networks (RNNs)~\cite{xu2019recurrent, hidasi2018recurrent} such as GRU4Rec~\cite{hidasi2015session} have been developed to model the interactions among adjacent items in a session. To further enhance the communications between non-adjacent items, transformers (TFs)~\cite{de2021transformers4rec, sun2019bert4rec} such as SASRec~\cite{kang2018self} adopt a self-attention to allow information exchange between non-adjacent items in a session. To capture even more complex transitional patterns, graph-based methods such as SR-GNN~\cite{wu2019session} extract the session graph for each session and use a gated graph neural network to learn session embeddings. Different from previous works that only capture transitional patterns within the session, we construct a global graph with different types of edges to capture even more broad information, the related works of which are reviewed next.

\subsubsection{Graph-based Recommendation System}
Graph, as a general data structure representing relations of entities, has been widely adopted to assist many real-world applications~\cite{graphacc, wang2023knowledge, zhao2018fair, liu2023interpretable} and one of the most representative examples is session recommendation. Previous works explore global transitional patterns across different sessions by querying the global item graph~\cite{zhang2021learning, huang2021graph, qiu2020exploiting}. \cite{huang2021graph} designs a global context-enhanced inter-session relation encoder to capture the inter-session item-wise dependencies. \cite{zhang2021learning} constructs the dual session graph to model the pair-wise transition relationship between items based on the global connections. \cite{qiu2020exploiting} constructs the global graph by merging all individual session graphs. The very recent work KSTT~\cite{zhang2021knowledge} resorts to an item-category knowledge graph for session recommendation. However, the proposed models in all the above works learn item embeddings from the global graph without any session-tailored modification. Only GCE-GNN~\cite{wang2020global} and GCARM~\cite{pan2021graph} consider session adaptation in aggregating neighbors' information. However, GCE-GNN quantifies the importance of neighbors based on their similarity to the whole session without differentiating central items. GCARM treats all transitions similarly without distinguishing different types of interactions. To handle these two issues, we design a session-adaptive propagation to query neighbors based on session contexts and interaction types, the effectiveness of which is verified in Section~\ref{sec-casestudy}. Note that in this work, although three types of item-item co-interaction edges are considered, the constructed knowledge graph has only one type of node, the item. We leave the inclusion of different node types as one future work, such as adding user nodes, which could provide a way to personalize the session recommendations.


\end{document}